\documentclass{article}
\usepackage{spconf,amsmath,graphicx,tikz,textcomp}
\usepackage[sort]{cite}
\usepackage[hidelinks]{hyperref}
\usepackage{booktabs}
\usepackage{flushend}
\usepackage{amssymb}
\usetikzlibrary{shapes,arrows}
\usepackage{psfrag}

\makeatletter
\def\blfootnote{\xdef\@thefnmark{}\@footnotetext}
\makeatother

\tikzstyle{block} = [draw, rectangle, 
    minimum height=3em, minimum width=5em]
\tikzstyle{block2} = [draw, rectangle, 
    minimum height=3em, minimum width=3em]
\tikzstyle{sum} = [draw, circle, node distance=1cm]
\tikzstyle{input} = [coordinate]
\tikzstyle{input2} = [coordinate]
\tikzstyle{output} = [coordinate]
\tikzstyle{pinstyle} = [pin edge={to-,thin,black}]

\title{A Hybrid Approach for Low-Complexity \\ Joint Acoustic Echo and Noise Reduction}
\name{Shrishti Saha Shetu, Naveen Kumar Desiraju, Miguel Martinez, Emanu\"{e}l A. P. Habets, Edwin Mabande}
\address{Fraunhofer IIS, Am Wolfsmantel 33, 91058 Erlangen, Germany \\
\small \textit{\{shrishti.saha.shetu, naveen.kumar.desiraju, miguel.martinez, emanuel.habets, edwin.mabande\}@iis.fraunhofer.de}}
\begin{document}
\maketitle
\begin{abstract}
Deep learning-based methods that jointly perform the task of acoustic echo and noise reduction (AENR) often require high memory and computational resources, making them unsuitable for real-time deployment on low-resource platforms such as embedded devices. We propose a low-complexity hybrid approach for joint AENR by employing a single model to suppress both residual echo and noise components. Specifically, we integrate the state-of-the-art (SOTA) ULCNet model, which was originally proposed to achieve ultra-low complexity noise suppression, in a hybrid system and train it for joint AENR. We show that the proposed approach achieves better echo reduction and comparable noise reduction performance with much lower computational complexity and memory requirements than all considered SOTA methods, at the cost of slight degradation in speech quality.

\end{abstract}

\section{Introduction}
\label{sec:intro}

Acoustic echo and noise reduction (AENR) is a highly desirable technology in communication devices, which aims to produce an estimate of the near-end speech signal by suppressing the undesired echo and background noise components captured by the microphone. In recent years, there has been a considerable increase in the use of deep neural network (DNN)-based approaches to achieve better acoustic echo reduction (AER) and AENR performance, either using hybrid approaches in combination with an adaptive filter \cite{peng2021acoustic, zhang2023two, mack2023hybrid, franzen2022deep, halimeh2020efficient, valin2021low}  or using end-to-end approaches \cite{braun2022task, indenbom2022deep, indenbom2023deepvqe, zhang2022multi, westhausen2021acoustic}. However, most of the state-of-the-art (SOTA) approaches have high memory and computational complexity requirements, making them unsuitable for deployment on resource-constrained platforms such as embedded devices. The main motivation for this study is to develop a high-quality solution for joint AENR with low computational complexity and memory requirements such that it can be integrated into embedded devices for real-time applications.

A number of hybrid and end-to-end approaches have been proposed in the literature for AER and joint AENR. In Peng et al. \cite{peng2021acoustic}, a hybrid approach was used for AER, where the error signal obtained at the output of an adaptive filter and the far-end signal were fed as inputs to a DNN post-filter. In Zhang et al. \cite{zhang2023two}, a similar approach was also followed for the task of AER. In \cite{mack2023hybrid}, only the error signal, as well as the echo estimate, were fed as inputs to a DNN post-filter as part of a hybrid approach for AENR. 
In \cite{braun2022task}, a two-stage DNN model was used as part of an end-to-end approach for AENR by splitting the tasks of AER and noise reduction (NR). In the first stage, a DNN was fed with the microphone and time-aligned far-end signals as input for performing the task of AER, while in the second stage, a second DNN was fed with the error signal as well as the echo estimate to perform the task of NR. Both DNNs were trained by minimizing the same cost function. In the Align-CRUSE method \cite{indenbom2022deep}, an end-to-end approach was used for AER, where a small DNN was first used to time-align the far-end signal with the microphone signal, and then both these signals were fed into the main DNN to yield a magnitude mask for the desired near-end speech signal. In their follow-up work, i.e., the  Deep-VQE method \cite{indenbom2023deepvqe}, the authors used a similar end-to-end approach, with an improved time-alignment method for joint AENR and dereverberation tasks. 

In recent years, numerous low complexity methods have been proposed for the NR task, which achieved SOTA performance in terms of NR and speech quality \cite{shetu2024ultra, schroter2022deepfilternet, schroter2022deepfilternet2, westhausen2020dual, choi2021real}. However, it has not been investigated comprehensively if the same architectures can be used for AER and AENR tasks.  So, in this paper, we propose to leverage the ULCNet model \cite{shetu2024ultra}, which showed good NR performance at ultra-low complexity, for the task of AENR by making appropriate modifications to the model. In particular, we propose a hybrid system combining a diagonalized partioned-block-frequency-domain-adaptive-Kalman-filter, as detailed in \cite{kuech2014state} and hereafter referred to as KF, with a DNN post-filter based on the ULCNet model. The modified ULCNet model takes three inputs, namely the far-end signal, the error signal, and the echo estimate. Our motivation for using a hybrid approach is that using the KF as a pre-processor for AER lightens the overall load for the DNN post-filter, enabling us to achieve good AENR performance with an ultra-low complexity model. We demonstrate that the ULCNet model can be successfully modified to perform the joint AENR and delivers better performance than having dedicated AER and NR models in series at almost half the computational cost.

This paper is structured as follows. In Section~\ref{sec:Proposed}, we explain the proposed processing method, including the modifications w.r.t. the original model in \cite{shetu2024ultra} to achieve joint AENR. Subsequently, we present the experiments, results, and discussions in Section~\ref{sec:Exp}, followed by the conclusions in Section~\ref{sec:Con}.

\section{Signal Model and Proposed Method}
\label{sec:Proposed}

\begin{figure}[t]
    \includegraphics[scale=0.56]{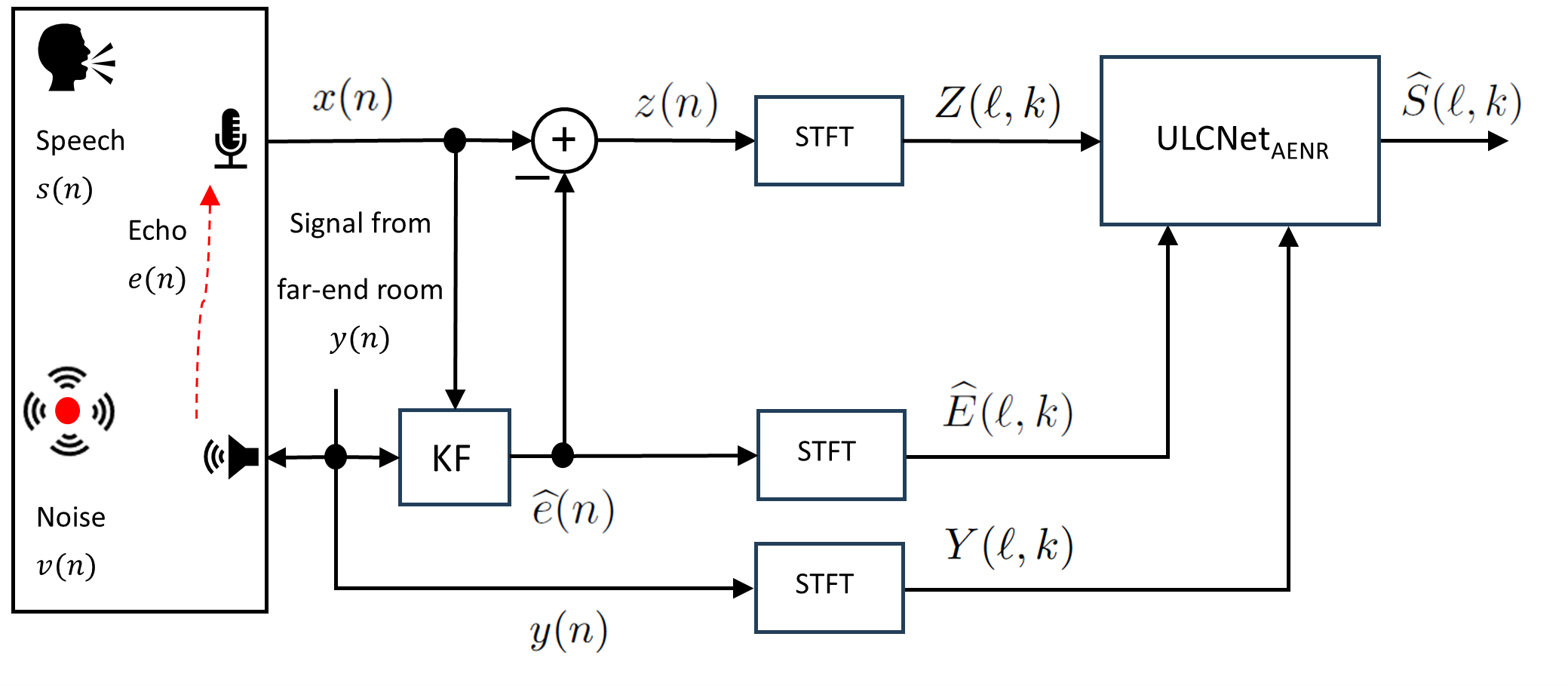}
    \caption{ Flow-diagram of proposed method}
    \label{Fig:SignalModel}
    \vspace{-0.3cm}
\end{figure}
Fig.\! \ref{Fig:SignalModel} shows a typical communication scenario where the far-end signal $y$ is played through the loudspeaker, and the microphone captures the acoustic echo $e$, the (desired) near-end speech $s$, and the background noise $v$. The microphone signal can thus be written as:
\begin{equation}
    x(n) = s(n) + e(n) + v(n),
    \label{eq:micTime}
\end{equation}
where $n$ denotes the discrete-time sample index. 

In order to suppress the echo and noise components, we propose a hybrid system consisting of two processing stages. In the first processing stage, a KF \cite{kuech2014state} generates an estimate for the echo signal, which is subtracted from the microphone signal to obtain the error signal:
\begin{equation*} 
    \begin{split} 
        z(n)    & = x(n) - \widehat e(n) \\
                & = s(n) + \Big\{ e(n) - \widehat e(n) \Big\} + v(n) \\
                & = s(n) + r(n) + v(n),
    \end{split}
    \tag{2}
    \label{eq:errorFreq}
    \vspace{-0.3cm}
\end{equation*}
where $z$, $\widehat{e}$, and $r$ denote the error signal, the echo estimate, and the residual echo signal, respectively. The residual echo is assumed to be composed of early residual echo due to filter misalignment, late residual echo due to reverberation, and non-linear echo components \cite{habets2008}.

\begin{figure}[t]
    \includegraphics[scale=0.84]{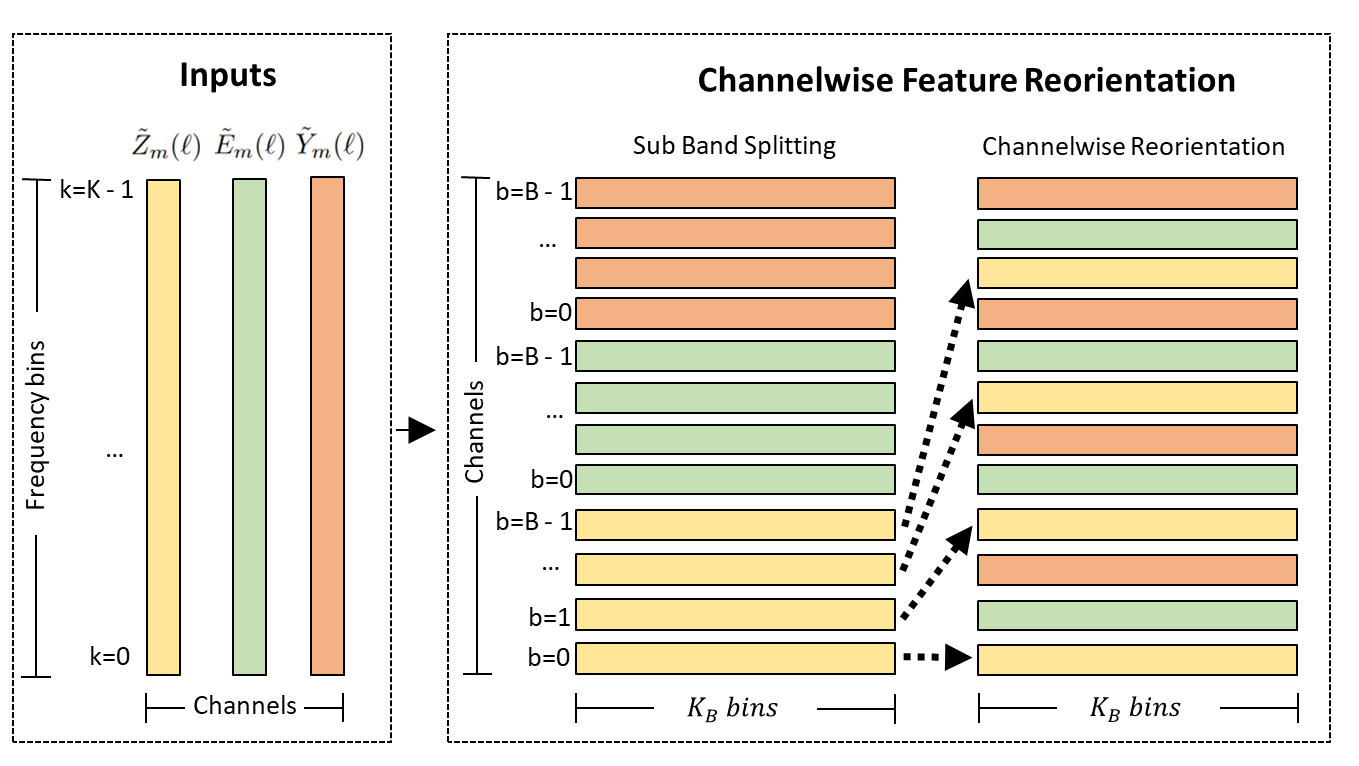}
    \caption{Modified channel-wise feature reorientation}
    \label{Fig:reorient}
    \vspace{-0.3cm}
\end{figure}

\textbf{\begin{table*}
\centering
\small
\begin{tabular} {l c c c c c c c c c c }
    \toprule
    \multicolumn{1}{l}{}                                & \multicolumn{2}{c}{Computational}     & \multicolumn{4}{c}{Interspeech 2021 \cite{cutler2021interspeech}} & \multicolumn{4}{c}{ICASSP 2023 \cite{cutler2024icassp}} \\
    \cmidrule(lr){4-7} \cmidrule(lr){8-11}
                                                        & \multicolumn{2}{c}{Complexity}        & \multicolumn{2}{c}{DT}        & FST           & NST               & \multicolumn{2}{c}{DT}        & FST           & NST           \\
    \cmidrule(lr){2-3} \cmidrule(lr){4-5} \cmidrule(lr){6-6} \cmidrule(lr){7-7} \cmidrule(lr){8-9} \cmidrule(lr){10-10} \cmidrule(lr){11-11}
    \textbf{Processing}                                 & Params [M]        & GMACs             & EMOS          & DMOS          & EMOS          & DMOS              & EMOS          & DMOS          & EMOS          & DMOS          \\
    \midrule
    Peng et al. \cite{peng2021acoustic}                 & 10.20             & 2.52              & 4.36          & 4.23          & 4.34          & 4.26              & -             & -	            & -             & -             \\
    Zhang et. al. \cite{zhang2023two}                   & 9.56              & -                 & -             & -             & -             & -                 & \textbf{4.72} & 4.16          & 4.70          & 3.91          \\
    Mack et al. \cite{mack2023hybrid}                   & 2.73              & 2.74              & 4.56          & 4.09          & \textbf{4.81} & 4.13              & -             & -	            & -             & -             \\
    Braun et al. \cite{braun2022task}                   & -                 & -                 & 4.55          & \textbf{4.25} & 4.35          & 4.18              & -             & -	            & -             & -             \\
    Align-CRUSE \cite{indenbom2022deep}                 & 0.74              & -                 & 4.45          & 4.07          & 4.67          & -                 & 4.60             & 3.95            & 4.56            & -             \\
    Deep-VQE \cite{indenbom2023deepvqe}                 & 7.50              & 4.02              & -             & -             & -             & -                 & 4.70          & \textbf{4.29} & 4.69          & \textbf{4.41} \\
    \hline
    KF                                                  & -                 & -                 & 3.02          & 3.77          & 3.02          & 4.16              & 2.75          & 3.51          & 3.38          & 4.07          \\
    ULCNet$_{\text{AER}}$                               & \textbf{0.69}     & \textbf{0.10}     & 4.40          & 3.82          & 4.39          & 4.17              & 4.31          & 3.31          & 4.65          & 4.06          \\
    ULCNet$_{\text{AER}}$ + ULCNet$_{\text{Freq}}$      & 1.38              & 0.20              & 4.53          & 3.80          & 4.46          & \textbf{4.35}     & 4.49          & 3.37          & 4.72          & 4.28          \\
    \textbf{Proposed ULCNet$_{\text{AENR}}$}            & \textbf{0.69}     & \textbf{0.10}     & \textbf{4.61} & 3.79          & 4.64          & 4.28              & 4.54          & 3.58          & \textbf{4.73} & 4.15          \\
    \bottomrule
\end{tabular}
\label{tab:Objectiveresult}
\caption{AECMOS \cite{purin2022aecmos} results on AEC Challenge blind test sets}
\vspace{-0.3cm}
\end{table*}}

In the second processing stage, the residual echo and noise components are jointly suppressed in the short-time Fourier transform (STFT) domain using a DNN post-filter. The signals $z(n)$, $\widehat e(n)$ and $y(n)$ are transformed into the STFT domain using a fast Fourier transform (FFT) of order $N_{\text{FFT}}$, with their STFT counterparts denoted as $Z(\ell,k)$, $\widehat{E}(\ell,k)$ and $Y(\ell,k)$, respectively. Here, $\ell$ denotes the frame index, $k$ denotes the frequency bin index, and $K = \frac{N_{\text{FFT}}}{2}+1$ the total number of frequency bins. As mentioned earlier, the proposed DNN post-filter is based on the ULCNet model from \cite{shetu2024ultra}, with three significant modifications to make it suitable for the joint AENR task:

\begin{enumerate}
    \item The proposed DNN post-filter takes three input signals, namely: \{$Z$, $\widehat{E}$, $Y$\} (instead of a single input signal \{$X$\} in \cite{shetu2024ultra}). The power-law compression method with a compression factor of $\alpha$ (as explained in Section~2.1 in \cite{shetu2024ultra}) is applied on all three inputs to obtain their respective compressed magnitudes \{$\widetilde{Z}_\textrm{m}$, $\widetilde{E}_\textrm{m}$, $\widetilde{Y}_\textrm{m}$\}. The left side of Fig.\! \ref{Fig:reorient} shows the compressed magnitudes for frame $\ell$.
    
    \item We propose a modified channel-wise feature reorientation and stacking method for our multiple compressed magnitude inputs, as shown in Fig.\! \ref{Fig:reorient}. Firstly, in each frame, we split the compressed magnitude features of each input into $B$ sub-bands of length $K_B$ frequency bins each, with an overlap factor between the sub-bands of $0 \leq \beta < 1$. Secondly, we interleave the resulting sub-bands of the three inputs as follows: $[\widetilde{Z}_{\textrm{m},0}~, \widetilde{E}_{\textrm{m},0}~, \widetilde{Y}_{\textrm{m},0} ~ \ldots ~ \widetilde{Z}_{\textrm{m},B-1}~, \widetilde{E}_{\textrm{m},B-1}~, \widetilde{Y}_{\textrm{m},B-1}~]$, where $\widetilde{Z}_{\textrm{m},b}$ denotes the $b^{\text{th}}$ sub-band obtained after splitting $\widetilde{Z}_\textrm{m}$. The interleaved features are then stacked together, as shown on the right side of Fig.~\ref{Fig:reorient}.

    \item The Intermediate Feature Computation block of ULCNet \cite{shetu2024ultra} now takes the phase component of the error signal $\widetilde{Z}_\textrm{p}$ as input (instead of the phase of the microphone signal $\widetilde{X}_\textrm{p}$). The compressed near-end speech estimate $\widetilde{S}$ is then computed using the complex ratio mask-based multiplication method shown in \cite{hu2020dccrn}, as follows:
    \begin{equation}
        \widetilde{S}(\ell, k) = \widetilde{Z}_\textrm{m}(\ell, k) \cdot M_\textrm{m}(\ell, k) \cdot e^{(\widetilde{Z}_\textrm{p}(\ell, k) + M_\textrm{p}(\ell, k))},
        \label{eq:CleanSpeechEst}
        \tag{3}
    \end{equation}
    where $M_\textrm{m}$ and $M_\textrm{p}$ represent the magnitude and phase components of the complex-valued mask $M$ computed by ULCNet, respectively. Finally, we obtain the near-end speech estimate $\widehat{S}$ by performing power-law decompression on the compressed near-end speech estimate $\widetilde{S}$, as described in Section~2.2 in \cite{shetu2024ultra}.
\end{enumerate}

\vspace{-0.25cm}
\section{Experiments and Results}
\label{sec:Exp}

\begin{table*}
\centering
\begin{tabular} {l c c c c c c}
    \toprule
    \multicolumn{1}{l}{}                                & \multicolumn{2}{c}{Computational Complexity}      & \multicolumn{4}{c}{DNS Challenge 2020 \cite{reddy2020interspeech}}\\
    \cmidrule(lr){2-3} \cmidrule(lr){4-7}
    \textbf{Processing}                                 & Params [M]        & GMACS                         & PESQ          & SI-SDR            & SIGMOS        & BAKMOS        \\
    \midrule
    Noisy                                               & -                 & -                             & 1.58          & 9.06              & 3.39          & 2.62          \\
    DeepFilterNet \cite{schroter2022deepfilternet}      & 1.78              & 0.35                          & 2.50          & 16.17             & 3.49          & 4.03          \\
    DeepFilterNet2 \cite{schroter2022deepfilternet2}    & 2.31              & 0.36                          & \textbf{2.65} & 16.60             & \textbf{3.51} & \textbf{4.12} \\
    ULCNet$_{\text{MS}}$\cite{shetu2024ultra}           & \textbf{0.68}     & \textbf{0.09}                 & 2.64          & 16.34             & 3.46          & 4.06          \\
    ULCNet$_{\text{Freq}}$\cite{shetu2024ultra}         & \textbf{0.68}     & \textbf{0.09}                 & 2.24          & \textbf{16.67}    & 3.38          & 4.09          \\
    ULCNet$_{\text{AER}}$ + ULCNet$_{\text{Freq}}$ & 1.38              & 0.20                          & 2.23          & 16.56             & 3.344         & 4.08          \\
    \textbf{Proposed ULCNet$_{\text{AENR}}$}       & 0.69              & 0.10                          & 2.11          & 15.58             & 3.302         & 4.05          \\
    \bottomrule
\end{tabular}
\label{tab:ObjectiveresultNR}
\caption{DNSMOS results on DNS Challenge 2020 \cite{reddy2020interspeech} synthetic non-reverb test set}
\vspace{-0.3cm}
\end{table*}

\subsection{Experimental Design}
\label{sec:training}

\noindent \textbf{Training Dataset}: We trained our proposed DNN post-filter for the task of AER-only as well as joint AENR. To create the training dataset for the task of AER, we used the synthetic and measured echo and far-end files provided in \cite{cutler2024icassp}. We manually curated the measured echo files and excluded files with uncharacteristically long echo delays (above 1.5s), as well as mismatched echo and far-end files. As the near-end signal, we used the clean and noisy speech signals provided in  \cite{cutler2024icassp, reddy2020interspeech}. To simulate the microphone signal and near-end training target, we mixed the signals as described in \cite{peng2021acoustic}, with a signal-to-echo ratio (SER) in the range [-20, 20] dB.

To create the training dataset for the joint AENR task, we used the clean speech and noise signals from the Interspeech 2020 DNS Challenge dataset \cite{reddy2020interspeech}. We first created the noisy mixtures with a signal-to-noise ratio (SNR) in the range $[-5, 30]$~dB, as described in \cite{shetu2024ultra}. Then we simulated different scenarios for the microphone signal (e.g., near-end single-talk (NST), far-end single-talk (FST), and double-talk (DT)) following the methods proposed in \cite{mack2023hybrid}, with an SER in the range $[-20,20]$ dB.

In total, we generated 1100 hours of training data each for both the AER and AENR tasks. All training samples were generated for a sampling frequency of 16 kHz.\\

\vspace{-0.2cm}
\noindent \textbf{Experimental Parameters}: For implementing the KF \cite{kuech2014state}, we derived the observation-noise power-spectral-density (PSD) matrix by recursively averaging the power of the error signal $Z$. The process-noise PSD matrix was estimated following the methods outlined in \cite{kuech2014state,haubner2021synergistic}. The Kalman gain used for the recursive averaging was set to 0.8. During training and inference, we always used 10 partitions.

As mentioned in Section~\ref{sec:Proposed}, we used the ULCNet model with the exact same model configurations as defined in \cite{shetu2024ultra} for both the AER and AENR tasks. As mentioned previously, we trained the ULCNet model always in combination with the KF and used the frequency domain mean-squared-error (MSE) loss function defined in \cite{shetu2024ultra}. From hereon, we denote the ULCNet model trained for the single-stage joint AENR task as $\text{ULCNet}_{\text{AENR}}$, and the model trained for only the AER task as $\text{ULCNet}_{\text{AER}}$. We use the pre-trained $\text{ULCNet}_{\text{Freq}}$ model from \cite{shetu2024ultra} as a baseline for the NR task and also to post-process the output of $\text{ULCNet}_{\text{AER}}$, such that it can be compared with the output of $\text{ULCNet}_{\text{AENR}}$.

To train both the $\text{ULCNet}_{\text{AER}}$ and $\text{ULCNet}_{\text{AENR}}$ models, we used the Adam optimizer with an initial learning rate of 0.004, which decayed by a factor of 10 when the validation loss did not improve with a patience of one epoch. Each training sample was of 3s duration, a batch size of 64 was chosen, and each model was trained for 20k steps per epoch. We chose $N_{\text{FFT}}=512$, such that $K = 257$, and a power-law compression factor $\alpha=0.3$. For channel-wise feature reorientation, we used $K_B=48$ with an overlap factor of $\beta=0.33$, such that $B = 8$.\\

\vspace{-0.2cm}
\noindent \textbf{Evaluation Dataset and Metrics}: For both the AER and AENR tasks, we use the blind test sets from the Interspeech 2021 \cite{cutler2021interspeech} and ICASSP 2023 \cite{cutler2024icassp} AEC Challenges for evaluation, which contain real-world recordings in diverse scenarios. To evaluate the AER performance of our proposed method in different scenarios and to compare it with SOTA methods, we use the AECMOS metrics \cite{purin2022aecmos}, which are composed of the DMOS and EMOS metrics, which measure the speech quality of the near-end speech estimate and the echo reduction performance, respectively. Additionally, to evaluate the NR performance, we compute the PESQ \cite{rix2001perceptual}, SI-SDR \cite{le2019sdr}, BAKMOS and SIGMOS metrics \cite{reddy2022dnsmos} on the DNS challenge 2020 synthetic non-reverb test set \cite{reddy2020interspeech}.

\subsection{Results and Discussion}
\label{sec:RD}

\noindent \textbf{AER Performance:}  We evaluate our proposed ULCNet$_{\text{AENR}}$ method against six existing SOTA methods from literature, namely Peng et al.\cite{peng2021acoustic}, Zhang et al.\cite{zhang2023two}, Mack et al.\cite{mack2023hybrid}, Braun et al.\cite{braun2022task}, Align-CRUSE \cite{indenbom2022deep} and Deep-VQE \cite{indenbom2023deepvqe}, as well as the KF output, the AER-only ULCNet$_{\text{AER}}$ method and the two-stage ULCNet$_{\text{AER}}$ + ULCNet$_{\text{Freq}}$ approach. We can observe from the results presented in Table 1 that the proposed ULCNet$_{\text{AENR}}$ method outperforms all other methods in terms of EMOS for the DT scenarios for the Interspeech 2021 test set and EMOS for the FST scenarios for the ICASSP 2023 test set. For both the DT and FST scenarios for either test set, all of our ULCNet-based approaches achieve comparable performance to SOTA methods in terms of the EMOS metrics, with ULCNet$_{\text{AENR}}$ performing best. For the NST scenarios for either test sets, the proposed ULCNet$_{\text{AENR}}$ method achieves comparable performance to SOTA methods in terms of the DMOS metrics, while the two-stage approach achieves better performance albeit at double the computational cost. However, all of our ULCNet-based approaches perform poorly as compared to SOTA methods in terms of the DMOS metrics for the DT scenarios. Please note that the proposed ULCNet$_{\text{AENR}}$ model is much smaller in size (up to 10x smaller) and computationally much cheaper (up to 4x cheaper) as compared to SOTA methods. \\

\vspace{-0.2cm}
\noindent \textbf{NR Performance:}  To evaluate our proposed method for the NR task, we compare with four different low-complexity SOTA methods, namely DeepFilterNet \cite{schroter2022deepfilternet}, DeepFilterNet2 \cite{schroter2022deepfilternet2}, ULCNet$_{\text{MS}}$ and ULCNet$_{\text{Freq}}$ \cite{shetu2024ultra}. We can observe from the results presented in Table 2 that the DeepFilterNet2 model outperforms all other methods while being the most computationally intensive. Our two-stage approach ULCNet$_{\text{AER}}$ + ULCNet$_{\text{Freq}}$ performs similarly to ULCNet$_{\text{Freq}}$, which was shown in \cite{shetu2024ultra} to achieve perceptually similar performance as DeepFilterNet2. We also observed similar perceptual quality in our informal listening tests for the two-stage approach. The ULCNet$_{\text{AENR}}$ model lags behind in all the objective metrics. However, this model was trained on a different training dataset as compared to the ULCNet$_{\text{Freq}}$ model, as it is designed for the joint AENR task. In informal listening tests, we found the perceptual quality of the joint AENR model to be comparable to the other methods. The processed samples can be found here: \url{https://fhgainr.github.io/fhgaenr/}. \\

\vspace{-0.2cm}
\noindent \textbf{Discussion:} Our proposed ULCNet$_{\text{AENR}}$ method, despite being computationally highly inexpensive, achieves comparable or better results as compared to SOTA approaches for the AER task across all scenarios and test sets, except for the DT scenarios in terms of the DMOS metric, which can be explained due to the aggressive nature of our method, combined with the effect of the modified power-law compression as discussed in \cite{shetu2024ultra}. One other important thing to note is that our proposed approach does not use any time-alignment method for the far-end signal, unlike most SOTA approaches \cite{braun2022task, indenbom2023deepvqe, indenbom2022deep}, which has shown to improve performance significantly in DT scenarios \cite{franzen2022deep}. We assume that combining a DNN-based time-alignment method with our proposed approach will further improve its performance in DT scenarios. We also observe that while being trained for the two separate tasks of AER and NR jointly, our ULCNet$_{\text{AENR}}$ method still achieves acceptable performance for the NR task. The low complexity of our proposed method makes it suitable for deployment on embedded devices, as it can run with a real-time factor of 13.1$\%$ on a Cortex-A53 1.43 GHz processor.

\vspace{-0.2cm}
\section{Conclusion}
\label{sec:Con}
\vspace{-0.2cm}
We proposed a low-complexity hybrid approach for joint AENR. Our proposed ULCNet$_{\text{AENR}}$ model achieves objective results on par with SOTA approaches, for both the AER and NR tasks, requiring by far the lowest computational complexity and model size, which makes it suitable for deployment in resource-constrained consumer devices.

\blfootnote{This work has been supported by the Free State of Bavaria in the DSAI project.}

\let\oldbibliography\thebibliography
\renewcommand{\thebibliography}[1]{%
  \oldbibliography{#1}%
  \footnotesize
  \setlength{\itemsep}{0pt}%
}
\bibliographystyle{IEEEbib}
\bibliography{strings}

\end{document}